\documentclass[a4]{article}
\usepackage{graphicx}
\usepackage{amssymb}
\usepackage{amsmath}
\usepackage{lscape}
\usepackage{mathrsfs}
\usepackage{textcomp}
\usepackage{epsfig}
\begin{document}
\begin{titlepage}
\begin{center}
{\Large {\bf 
%\sc
{Friedel sum rule in the presence of topological defects for graphene
}}} 
\\[1cm]

Baishali Chakraborty$^{a}, $\footnote{baishali.chakraborty@saha.ac.in} 
\hspace{.2cm} 
Kumar S. Gupta$^{a}, $\footnote{kumars.gupta@saha.ac.in}
\hspace{.2cm} 
Siddhartha Sen$^{b}, $\footnote{siddhartha.sen@tcd.ie}
\\[0.5cm]

${}^a$ Theory Division, Saha Institute of Nuclear Physics, 1/AF Bidhannagar, Calcutta 700064, India
\\[0.5cm]

${}^b$ CRANN, Trinity College Dublin, Dublin 2, Ireland
\end{center}
\vspace{1cm}

\begin{abstract}
\noindent
The Friedel sum rule is extended to deal with topological defects for the case of a graphene cone 
in the presence of an external Coulomb charge. The dependence in the way the number of states change 
due to both the topological defect as well as the Coulomb charge are studied. Our analysis addresses 
both the cases of a subcritical as well as a supercritical value of the Coulomb charge. We also discuss 
the experimental implications of introducing a self-adjoint extension of the system Hamiltonian. We argue that the 
boundary conditions following from the self-adjoint extension encode the effect of short range interactions present 
in the system. 
\end{abstract}
\end{titlepage}
%\pacs{81.05.ue, 03.65.Ge, 71.30.+h}
% \maketitle

\section{Introduction}
The Friedel sum rule provides a method for getting information about 
polarization charge due to an external charge impurity in the system\cite{friedel, mahan}. 
The screening charge around the impurity is
directly proportional to the change in the number of states
$\Delta N$ due to the Coulomb potential and Friedel sum rule 
shows that $\Delta N$ can be expressed in terms of a summation of
scattering phase shifts at Fermi energy for all the angular momentum 
channels\cite{lin1,lin2,lin3, moroz1, moroz2}.
As $\Delta N$ is related to the LDOS of the system,
the properties of the system which are related to LDOS
can be obtained using this rule. 

In this paper we analyze the Friedel sum rule in a gapless graphene
cone in the presence of an external Coulomb charge\cite{critical1}.
The strength of the external Coulomb charge introduced  in graphene can be classified
as either subcritical or supercritical. The critical value of the Coulomb charge
corresponds to a situation beyond which the system becomes quantum mechanically unstable\cite{critical1,castro,levi1,levi2,kats1,us1}
and this leads to the \textquotesingle fall to the centre\textquotesingle\cite{levi1,levi2} phenomenon. 
Graphene, experimentally fabricated in $2004$\cite{novo1,novo2,zhang},
provides an ideal laboratory to study this phenomenon. This is due
to the fact that the Dirac type quasiparticles in graphene 
have a Fermi velocity which is approximately $300$ times 
smaller than the velocity of light. Thus the supercriticality
is easily reached in graphene in presence of a relatively small external Coulomb charge 
impurity $Ze \sim 1$\cite{castro,levi1,levi2,kats1,us1} and
the atomic collapse\cite{levi2} in this region leads to the formation of quasibound states.
Recently such quasibound states have been observed experimentally\cite{levi3}
in gapless planer graphene. In this paper we study the supercritical
Coulomb impurity in graphene in presence of a conical defect.
The wavefunctions associated with the gapless Dirac type excitations in pristine graphene\cite{wall,mele,sem,geim,rmp1,rmp2,rmp3} 
pick up holonomy when the quasiparticles move around a closed path
encircling a conical defect\cite{crespi1,crespi2,stone1}. 
The holonomies due to topological defects\cite{crespi1,crespi2,stone1,voz1,voz2,osi1,
sitenko,voz3,voz4,mudry,furtado,stone2,voz5,Gonzalez,yazyev,fonseca,voz6,Guinea1,furtado2,voz7} can be realized 
by introducing a suitable flux tube passing through the origin\cite{Jackiw1,gerbert,yam,Jackiw2,Jackiw3,Jackiw4,Jackiw5}. 
We are thus lead to study the combined effect of the flux tube potential and the external Coulomb charge on 
graphene.

In addition to the supercritical region we analyze our system thoroughly
for the subcritical values of the Coulomb charge. In the subcritical region
for a certain range of system parameters we found that 
a single real parameter is required for labeling the  
boundary conditions at the location of the defects. To understand the physical origin of such a parameter we recall that 
the Coulomb charge as well as the conical defect can give rise to short range
interactions in graphene. Such interactions cannot be directly incorporated in the Dirac equation
as the latter is valid only in the low energy or long-wavelength limit. However
the combined effect of such short range interactions can be encoded in
the boundary conditions\cite{reed,falomir,critical2,ksg1} by the parameter.
This additional real parameter ensures current conservation leading to a 
self-adjoint Hamiltonian and unitary evolution of the quantum system\cite{dipti1,dipti2,dipti3}. 
We show that the scattering phase shift and consequently $\Delta N$ depend 
explicitly on this parameter labeling the boundary 
conditions in the graphene cone. This parameter is determined empirically 
as it cannot be obtained by theory.

This paper is organized as follows. In the next Section we set up 
the Friedel sum rule for gapless graphene cone with a point charge at the 
apex. Then the analysis of the spectrum is done in the subcritical region, 
where we obtain the scattering phase shifts and change in the number of states
and show how these physical quantities depend explicitly on the sample topology. 
After that we discuss the effect of generalized boundary conditions on the spectrum. 
In the next section the analysis of the corresponding spectrum is done in the supercritical region.
We end this paper with some discussion and outlook. 

\section{Friedel Sum Rule for massless graphene cone}
The low energy properties of the quasiparticles near a Dirac point in a planer gapless graphene
sample\cite{wall,mele,sem,geim,rmp1,rmp2,rmp3} in the presence of an external Coulomb charge is given by 
\begin{equation}
\label{f0}
\left [ -i\hbar v_F (\sigma_1 \partial_x + \sigma_2 \partial_y) + \sigma_0 \left ( \frac{ - \alpha}{r} \right ) 
\right ] \Psi = E \Psi, 
\end{equation}
where $r$ is the radial coordinate in the two-dimensional $x-y$ plane and $\alpha$ is 
the Coulomb interaction strength. The Pauli matrices $\sigma_{1,2,3}$ and the identity 
matrix $\sigma_0$ act on the pseudospin indices $A,B$.

When a conical defect is introduced in graphene by removing $n$ number of sectors
subtending an angle $\frac{2n\pi}{6}$ at the centre and the edges
of the removed sector are identified, 
the angular boundary condition obeyed by the Dirac spinor is modified. 
Due to the identification of the two edges of the removed sector the frame 
$\{\hat{e}_x,\hat{e}_y\}$ becomes discontinuous across 
the joining line. Therefore we choose a new set of frames which is rotated 
with respect to the old frame $(\hat{e}_x,\hat{e}_y)$ by an angle $\omega =\theta + \frac{\pi}{2}$ \cite{critical1}. 
%given by 
%\begin{eqnarray}
%\label{f0.1}
%\hat{e}_{x^{\prime}} = \hat{e}_\theta~~~~~~~~~~\mbox{and}~~~~~~~~~~~\hat{e}_{y^{\prime}} = -\hat{e}_r,
%\end{eqnarray}
The effect of the conical topology can be equivalently described by 
introducing a magnetic flux tube passing through the centre of the plane graphene sheet.
The magnetic vector potential associated with the flux tube replaces the ordinary derivatives in the Hamiltonian by 
the corresponding covariant derivatives . Therefore the Dirac equation for massless 
graphene cone becomes
\begin{eqnarray} 
\label{f1}
 H \Psi_{\nu} = E_{\nu} \Psi_{\nu},
\end{eqnarray}
where
\begin{eqnarray} 
\label{f2}
H  = \left( 
\begin{array}{cc}
-\frac{\alpha}{r} &  \partial_r - \frac{i}{r(1-\frac{n}{6})}\partial_\theta \pm \frac{\frac{n}{4}}{r(1-\frac{n}{6})} + \frac{1}{2r}  \\
-\partial_r - \frac{i}{r(1-\frac{n}{6})}\partial_\theta  \pm \frac{\frac{n}{4}}{r(1-\frac{n}{6})}-\frac{1}{2r} & 
 -\frac{\alpha}{r}
\end{array}
\right).\nonumber
\end{eqnarray}
For the wave function $\Psi_{\nu}$ we use the following ansatz.
\begin{eqnarray}
\label{f3}
\Psi_{\nu}=\sum_j \left( 
\begin{array}{c}
\Psi_{A\nu}^{(j)}(r)\\
\Psi_{B\nu}^{(j)}(r)
\end{array}
\right)e^{ij\theta},
\end{eqnarray}
where $j$ is the total angular momentum quantum number.
%$\beta=\sqrt{\nu^2 -\alpha^2}$ and
%\Psi_{A\lambda}^{(j)}(r)= v_\lambda + u_\lambda,~
%\Psi_{B\lambda}^{(j)}(r)= v_\lambda - u_\lambda$
The radial Dirac equation in each angular momentum channel $j$ is given by
\begin{eqnarray}
\label{f3.1}
\left( 
\begin{array}{cc}
E_{\nu}+\frac{\alpha}{r} & -\{ \partial_r +(\lambda +\frac{1}{2})\frac{1}{r}\}  \\
\{\partial_r -(\lambda - \frac{1}{2})\frac{1}{r}\} & E_{\nu}+\frac{\alpha}{r}
\end{array}
\right) \left( 
\begin{array}{c}
\Psi_{A\nu}^{(j)}(r)\\
\Psi_{B\nu}^{(j)}(r)
\end{array}
\right)
=0,
\end{eqnarray}
where $\lambda= \frac{j\pm \frac{n}{4}}{1-\frac{n}{6}}$. In the absence of the 
external Coulomb potential the equations for the components of $\Psi_{\nu}^j (r)$
can be decoupled into Bessel equations.
The equation for the Dirac spinor component $\Psi_{A\nu}^j(r)$ is given by 
\begin{eqnarray}
\label{f3.2a}
\frac{d^{2}}{d r^{2}} \Psi_{A\nu} + \frac{1}{r}\frac{d}{dr}\Psi_{A\nu} + \left[E_{\nu}^2 - \frac{(\lambda-\frac{1}{2})^2}{r^2}\right]\Psi_{A\nu}=0
\end{eqnarray}
%and 
%\begin{eqnarray}
%\label{f3.2b}
%\frac{d^{2} \Psi_{B\lambda}}{d r^{2}} + \frac{1}{r}\frac{d}{dr}\Psi_{B\lambda} + \left[E_{\lambda}^2 - \frac{(\nu+\frac{1}{2})^2}{r^2}\right]\Psi_{B\lambda}=0.
%\end{eqnarray}
and the solution regular at the origin is
 \begin{eqnarray}
\label{f3.3}
\Psi_{A\nu}^j(r)=C(E_{\nu}) J_{\lambda-\frac{1}{2}}(E_{\nu}r).
\end{eqnarray}
Using a suitable normalization condition we get the expression for $C(E_{\nu})$ to be $\sqrt{\frac{E_{\nu}}{2}}$ .
From $\Psi_{A\nu}$ the expression for $\Psi_{B\nu}$ can be obtained with the help of 
Eq.(\ref{f3.1}). When $r\rightarrow\infty$, the asymptotic expression for $\Psi_{A\nu}$
is given by 
\begin{eqnarray}
 \label{f3.4a}
\Psi_{A\nu}^j(r)\rightarrow \frac{1}{\sqrt{\pi r}} \mbox{cos}\left(E_{\nu}r - \frac{\lambda \pi}{2}\right).
\end{eqnarray}
Substituting this expression in Eq. (\ref{f3.1}) we get, 
\begin{eqnarray}
 \label{f3.4b}
\lim_{r\rightarrow\infty}\Psi_{B\nu}^j(r)\rightarrow \frac{1}{\sqrt{\pi r}} \mbox{sin}\left(E_{\nu}r - \frac{\lambda \pi}{2}\right).
\end{eqnarray}
To evaluate the total change in the number of states  $\Delta N$ around the Coulomb charge at the apex of 
the massless graphene cone, now we consider a two dimensional circular area of a large radius $R$.
This area has the Coulomb charge at its centre. The magnetic flux tube representing the 
nontrivial holonomies produced by the conical defect also passes through the centre. 
The asymptotic behaviors of the wave function will be used for the evaluation process.
We proceed with multiplying Eq. (\ref{f1}) by the adjoint of the Dirac spinor 
and the adjoint of Eq. (\ref{f1}) by the Dirac spinor.

The adjoint of Eq.(\ref{f1}) is given by
\begin{eqnarray} 
\label{f4}
(H \Psi_{\nu})^{\dagger} = E_{\nu} \Psi_{\nu}^{\dagger}.
\end{eqnarray}
Multiplying Eq.(\ref{f4}) by $\Psi_{\nu^\prime}$ and Eq. (\ref{f1}) for $\Psi_{\nu^\prime}$ 
by $\Psi_{\nu}^{\dagger}$ and subtracting we obtain the following .
\begin{eqnarray}
 \label{f5}
(E_{\nu^\prime}-E_{\nu})\Psi_{\nu}^{\dagger}\Psi_{\nu^\prime}=[-i\vec{\nabla}\cdotp\{\Psi_{\nu}^{\dagger}\vec{\sigma}\Psi_{\nu^\prime}\}]+\frac{i}{r}\{\Psi_{\nu}^{\dagger} \sigma_2 \Psi_{\nu^\prime}\}.
\end{eqnarray}

Now integrating Eq.(\ref{f5}) over the whole area we have
\begin{eqnarray}
\label{f6}
 (E_{\nu^\prime}-E_{\nu})\iint d^2 r \Psi_{\nu}^{\dagger}\Psi_{\nu^\prime} = \iint d^2 r [-i\vec{\nabla}\cdotp\{\Psi_{\nu}^{\dagger}\vec{\sigma}\Psi_{\nu^\prime}\}]+\iint d^2 r \frac{i}{r}\{\Psi_{\nu}^{\dagger} \sigma_2 \Psi_{\nu^\prime}\}.
\end{eqnarray}
Application of divergence theorem gives
\begin{eqnarray}
\label{f7}
 \iint d^2 r \Psi_{\nu}^{\dagger}\Psi_{\nu^\prime} = \frac{R}{(E_{\nu^\prime}-E_{\nu})}\oint d\theta (-i)\{\Psi_{\nu}^{\dagger}(\vec{\sigma}\cdotp \hat{r})\Psi_{\nu^\prime}\}+\iint \frac{d^2 r}{(E_{\nu^\prime}-E_{\nu})} \frac{i}{r}\{\Psi_{\nu}^{\dagger} \sigma_2 \Psi_{\nu^\prime}\}.
\end{eqnarray}
At large distance, the second term on the R.H.S. of Eq. (\ref{f7}) gives negligible contribution.
Therefore the above integral can be expanded as 
\begin{eqnarray}
\label{f8}
 \iint d^2 r \Psi_{\nu}^{\dagger}\Psi_{\nu^\prime} = \frac{2\pi R}{(E_{\nu^\prime}-E_{\nu})}\sum_j [\Psi_{A\nu}^{j*}(r)\Psi_{B\nu^\prime}^j(r)-\Psi_{B\nu}^{j*}(r)\Psi_{A\nu^\prime}^j(r)].
\end{eqnarray}
Eq. (\ref{f8}) gives the local density of states at a particular energy level $E_{\nu}$.
So the total change in the number of states around the Coulomb potential can be found
by integrating the expression up to the Fermi energy level $E_F$ in the presence
and in the absence of the external Coulomb potential and then by obtaining the difference
between the two integrals.
\begin{eqnarray}
 \label{f9}
\Delta N = \lim_{r\longrightarrow\infty}\lim_{E_{\nu^\prime}\longrightarrow E_{\nu}} \int_{0}^{E_F} dE_{\nu}\iint d^2 r [\Psi_{\nu}^{\dagger}\Psi_{\nu^\prime}-\Psi_{\nu 0}^{\dagger}\Psi_{\nu^\prime 0}].
\end{eqnarray}
Here $\Psi_{\nu 0}$ represents the Dirac spinor of the massless graphene cone 
in the absence of the external Coulomb potential. Now putting the asymptotic expression 
of the wave function in Eq. (\ref{f9}) we can obtain the Friedel sum rule for massless graphene 
cone \cite{lin3}. According to the rule 
\begin{eqnarray}
 \label{f10}
\Delta N = \frac{1}{\pi}\sum_j [\delta_j (E_F)-\delta_j (0)].
\end{eqnarray}
Here $\delta_j$ represents the scattering phase shift in the $j$-th 
angular momentum channel. As the scattering phase shift contains the 
term $\lambda$ both for the subcritical and supercritical region, the 
Friedel sum rule depends explicitly on the topological defect of the system.

This rule can also be established by calculating directly the DOS 
using the Green function $G_{\alpha}(\mathbf{r_1}, \mathbf{r_2}, E + i\epsilon)$  \cite{moroz1,moroz2}
and applying the formula
\begin{eqnarray}
 \label{fg1}
\rho_{\alpha}(E)= - \frac{1}{\pi} \mbox{Im Tr}G_{\alpha}(\mathbf{r_1}, \mathbf{r_2}, E + i\epsilon).
\end{eqnarray}
The Green function can be expanded in terms of the eigenfunctions $\Psi_{\nu}(\mathbf{r},\theta)$ in polar coordinates.
Using the eigenfunctions in the presence and in the absence of the Coulomb potential 
the change in DOS can be calculated. The change in the number of states can then
be obtained by  
\begin{eqnarray}
 \label{fg2}
\Delta N = \int_{-\infty}^{E_F} dE^{\prime} [\rho_{\alpha}(E^{\prime})- \rho_{0}(E^{\prime})].
\end{eqnarray}
Putting the expressions of the eigenfunctions in the Green function we can find out that
the change in number of states depends on the sum of the scattering phase shifts over all
the angular momentum channels.
 
We shall analyse the effect of conical topology on the Friedel sum rule
for massless graphene for both the subcritical and supercritical region
in the following two sections. 
%Our calculations will be restricted in the
%region where the Dirac Hamiltonian is self-adjoint.

\section{Subcritical region}

In this section we shall obtain the expression of scattering phase shift 
for the massless graphene with a conical defect in presence of a subcritical Coulomb charge.
To solve the Dirac equation Eq. (\ref{f1}) 
in presence of a subcritical Coulomb charge we assume
\begin{eqnarray}
\label{f11}
\Psi_{\nu}(r,\theta)=\sum_j \left( 
\begin{array}{c}
\Psi_{A\nu}^{(j)}(r)\\
i\Psi_{B\nu}^{(j)}(r)
\end{array}
\right)e^{-iE_{\nu}r} {r}^{\beta - (1/2)}e^{ij\theta},
\end{eqnarray}
where $\beta=\sqrt{\lambda^2 -\alpha^2}$ and we use two new functions
$u_\nu^{(j)}(r)$ and $v_\nu^{(j)}(r)$ defined by
$\Psi_{A\nu}^{(j)}(r)= v_\nu^{(j)}(r) + u_\nu^{(j)}(r),~
\Psi_{B\nu}^{(j)}(r)= v_\nu^{(j)}(r) - u_\nu^{(j)}(r)$
to get the following equations.
\begin{equation} \label{f12}
r \frac{d v_\nu^{(j)}(r)}{d r} + (\beta + i\alpha) v_\nu^{(j)}(r) - \lambda u_\nu^{(j)}(r) = 0~~~~~~~~~
\end{equation}
and 
\begin{equation} \label{f13}
r \frac{d u_\nu^{(j)}(r)}{d r} + (\beta - i \alpha -2iE_{\nu}r) u_\nu^{(j)}(r) - \lambda v_\nu^{(j)}(r) = 0.~~~~~~~~~
\end{equation}
Combining Eqs. (\ref{f12}) and (\ref{f13}) we get
\begin{equation} \label{f14}
 s \frac{d^{2} v^{(j)}(s) }{d s^{2}} + (1 + 2\beta - s)\frac{d v^{(j)}(s)}{d s} - \left ( \beta + i \alpha \right ) v^{(j)}(s) = 0,
\end{equation}
where $s= -2ikr$, with  $k=- E_{\nu}$.

The solution of Eq. (\ref{f14}) which is regular at the origin is given by
\begin{eqnarray}
 \label{f15}
v^{(j)}(s) = A M \left ( \beta + i\alpha,~ 1 + 2\beta,~s \right),
\end{eqnarray}
where $M$ is the confluent hypergeometric function\cite{stegun} and $A$ is a constant which 
depends on the energy of the system.

\begin{figure}
[ht] 
\centering
\includegraphics[bb= 280 14 12 220]{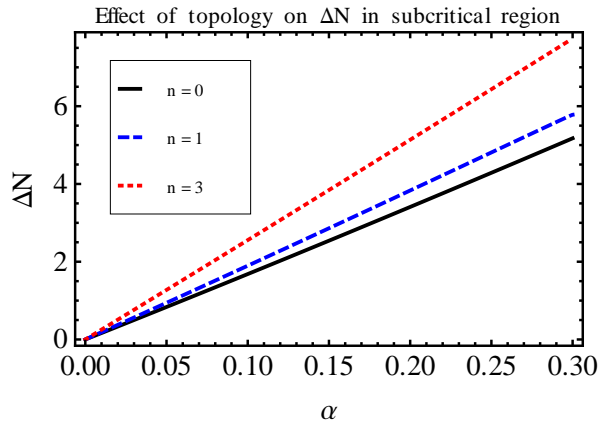}
 % f1.eps: 0x0 pixel, 0dpi, 0.00x0.00 cm, bb=240 14 212 40
\caption{Dependence of change in the number of states on subcritical Coulomb potential using Friedel sum rule.}
\label{fig:1}
\end{figure}

Substituting this expression of $v^{(j)}(s)$ from Eq.(\ref{f15})
in Eq.(\ref{f12}) we have 
\begin{eqnarray}
 \label{f16}
u^{(j)}(s) = A \frac{(\beta+i\alpha)}{\lambda} M \left ( 1+\beta + i\alpha,~ 1 + 2\beta,~s \right).
\end{eqnarray}
Using the asymptotic form of $v^{(j)}(s)$ and $u^{(j)}(s)$ we can find
out the expression of the scattering phase shift to be
\begin{eqnarray}
 \label{f17}
\delta_{j}(k)= -\alpha \ln(2kr) + \arg[\Gamma(1+\beta+i\alpha)] - \frac{\pi \beta}{2} - \frac{1}{2}\tan^{-1}\left(\frac{\alpha}{\beta}\right) + \left|\frac{\lambda \pi}{2}\right|.
\end{eqnarray}
With the help of this expression and the Friedel sum rule we have found
out the dependence of $\Delta N$ on the Coulomb potential and the conical 
defect in massless graphene. We have plotted the dependence in Fig.(\ref{fig:1}).
The change in the number of states is directly proportional to the polarization charge
induced in the system by the external Coulomb charge. Therefore from Fig.(\ref{fig:1})
we can determine the dependence of polarization charge on subcritical Coulomb potential using
Friedel sum rule.

From the plot we can see that for  different values of $n$ i.e. for different topology 
the polarization charge increases with the subcritical Coulomb potential at a different rate.
It indicates that the change in the number of states around the external subcritical Coulomb charge
depends on the topology of the system and with the increase in the angular deficit of the
cone the rate of this change increases.

\subsection{Generalized boundary conditions}

The conical defect and the Coulomb charge impurity can give rise to some 
short range interactions in the graphene system. These interactions cannot 
be included as dynamical terms in the Dirac equation
as the latter is valid for only low energy and long wavelength excitations.
However, through the choice of suitable boundary conditions prescribed by von Neumann
for systems with unitary time evolution and probability current conservation\cite{dipti1,dipti2,dipti3}, 
combined effect of those interactions can be considered 
which is discussed below\cite{reed,falomir,critical2,ksg1}.

The Dirac operator $H$ in Eq.(\ref{f1}) has an angular part and a radial part. 
The angular boundary condition is kept unchanged as the angular part operates 
on a domain spanned by the antiperiodic functions $e^{ij\theta}$ where $j$ is 
a half integer.
The radial Dirac operator $H_{r}$ is given by 
\begin{eqnarray}
\label{sae1}
H_{r}=
\left( 
\begin{array}{cc}
-\frac{\alpha}{r} & \{ \partial_r +(\lambda +\frac{1}{2})\frac{1}{r}\}  \\
-\{\partial_r -(\lambda - \frac{1}{2})\frac{1}{r}\} & -\frac{\alpha}{r}
\end{array}
\right).
\end{eqnarray}
It is symmetric in the domain $\mathcal{D}_{0} = C_{0}^{\infty}(R^{+})$ consisting of infinitely differentiable 
functions of compact support in the real half line $R^{+}$ and its adjoint 
operator $H_{r}^{\dagger}$ has the same expression as $H_{r}$ but its domain can be different.
The domain of self-adjointness of the operator $H_r$ can be determined by using the equation 
\begin{equation}\label{sae2}
 H_{r}^{\dagger}\Psi_{\pm}=\pm\frac{i}{l}\Psi_{\pm},
\end{equation}
where $l$ has the dimension of length and
\begin{eqnarray}
\label{sae3}
\Psi_{\pm}(r)=\sum_j \left( 
\begin{array}{c}
v_\pm^{(j)}(r) + u_\pm^{(j)}(r)\\
i(v_\pm^{(j)}(r) - u_\pm^{(j)}(r))
\end{array}
\right)e^{\pm \frac{r}{l}} r^{\beta - \frac{1}{2}}.
\end{eqnarray}
The total number of square integrable, linearly independent solutions of Equation(\ref{sae2})
gives the deficiency indices for $H_r$ and they are denoted by $n_{\pm}$.
%For obtaining $n_{\pm}$, Equation(\ref{f1}) is considered with $E$ replaced by $\pm\frac{i}{l}$.
The existence of imaginary eigenvalues $\pm\frac{i}{l}$ in the spectrum is a measure of the deviation of 
the operator $H_r$ from self-adjointness. The non zero deficiency indices serve as the measurement of this deviation.
The deficiency indices classify $H_r$ in three different ways \cite{reed} :
$(1)$ When $n_+ = n_- =0$, $H_{r}$ is essentially self-adjoint in $\mathcal{D}_{0}(H_{\rho})$. 
$(2)$ When $n_+ = n_-\neq0$, $H_{r}$ is not self-adjoint in $\mathcal{D}_{0}(H_{\rho})$ but it 
can admit self-adjoint extensions. 
$(3)$ When $n_+\neq n_-$, $H_{r}$ cannot have self-adjoint extensions. 

Eq. (\ref{sae2}) leads to the following two coupled differential equations:
\begin{equation} \label{sae4}
r \frac{d v_\pm^{(j)}(r)}{d r} + (\beta + i\alpha) v_\pm^{(j)}(r) - \lambda u_\pm^{(j)}(r) = 0~~~~~~~~~
\end{equation}
and 
\begin{equation} \label{sae5}
r \frac{d u_\pm^{(j)}(r)}{d r} + (\beta - i \alpha \pm \frac{2r}{l}) u_\pm^{(j)}(r) - \lambda v_\pm^{(j)}(r) = 0.~~~~~~~~~
\end{equation}
Combining Eqs. (\ref{sae4}) and (\ref{sae5}) we get
\begin{equation} \label{sae6}
 s \frac{d^{2} v_{\pm}^{(j)}(s) }{d s^{2}} + (1 + 2\beta - s)\frac{d v_{\pm}^{(j)}(s)}{d s} - \left ( \beta + i \alpha \right ) v_{\pm}^{(j)}(s) = 0,
\end{equation}
where $s= \mp\frac{2r}{l}$ for $E=\pm \frac{i}{l}$.

In order to solve Eq. (\ref{sae6}) let us first consider the case where $s=-\frac{2r}{l}$ i.e. $E=\frac{i}{l}$.
The solution can be written as
\begin{equation}\label{sae7}
v_{+}= e^{-\frac{2r}{l}} U\left(1+\beta-i\alpha,1+2\beta,\frac{2r}{l}\right).
\end{equation}
Putting this expression for $v_{+}$ in Eq. (\ref{sae4}) we obtain
\begin{equation}\label{sae8}
u_{+}= -\frac{e^{-\frac{2r}{l}}}{\lambda} U\left(\beta-i\alpha,1+2\beta,\frac{2r}{l}\right).
\end{equation}
Therefore the radial part of the upper component of the wave function becomes
\begin{equation}\label{sae9}
\Psi_{A+}= e^{-\frac{r}{l}}\left[U\left(1+\beta-i\alpha,1+2\beta,\frac{2r}{l}\right)-\frac{1}{\lambda} U\left(\beta-i\alpha,1+2\beta,\frac{2r}{l}\right)\right]r^{\beta-\frac{1}{2}}.
\end{equation}
This component $\Psi_{A+}$ spans the $E=+\frac{i}{l}$ deficiency subspace. In order to find 
conditions under which $\Psi_{A+}$ and consequently $\Psi_{+}$ has square integrable solutions,
we notice that as $r\rightarrow\infty,~\Psi_{A+}\rightarrow0$. As a result we can say $\Psi_{A+}$
is square integrable at infinity. When $r\rightarrow0$,
\begin{equation}\label{sae10}
\int |\Psi_{A+}|^2 r dr \sim \int r^{-2\beta} dr + \mbox{converging terms}.
\end{equation}
Therefore we can say that $\Psi_{A+}$ is a square integrable function for 
the range $0<\beta<\frac{1}{2}$. Proceeding in the similar manner we can 
show that for the specified range of $\beta$ the entire radial wave function 
is square integrable and the deficiency index $n_+ = 1$ for a graphene cone
in presence of an external Coulomb charge.

Next we consider the case where $s=\frac{2r}{l}$ i.e. $E=-\frac{i}{l}$.
In this case the solution can be written as
\begin{equation}\label{sae11}
v_{-}= U\left(\beta+i\alpha,1+2\beta,\frac{2r}{l}\right).
\end{equation}
Putting this expression for $v_{-}$ again in Eq. (\ref{sae4}) we obtain
\begin{equation}\label{sae12}
u_{-}= \frac{(\beta + i\alpha)(-\beta + i\alpha)}{\lambda} U\left(1+\beta+i\alpha, 1+2\beta, \frac{2r}{l}\right).
\end{equation}
Therefore we have 
\begin{equation}\label{sae13}
\Psi_{A-}= e^{-\frac{r}{l}}\left[U\left(\beta + i\alpha,1+2\beta,\frac{2r}{l}\right)+\frac{(\beta + i\alpha)(-\beta + i\alpha)}{\lambda} U\left(1+\beta+i\alpha,1+2\beta,\frac{2r}{l}\right)\right]r^{\beta-\frac{1}{2}}.
\end{equation}
as the radial part of the upper component of the wave function which spans the $E=-\frac{i}{l}$ deficiency subspace. 
Analysing as before, we notice that when  $0<\beta<\frac{1}{2}$, for $E=-\frac{i}{l}$ also we have a single square 
integrable solution for the wave-function indicating $n_-=1$.

As the deficiency indices $n_+ = n_- = 1$,  following von Neumann's analysis we can say that the radial Hamiltonian $H_r$
admits a one parameter family of self-adjoint extension in this case. The domain representing the boundary conditions
for which $H_r$ is self-adjoint is given by 
${\mathcal{D}}_{\Phi}(H_r) = {\mathcal{D}}_{0}(H_{r})\oplus  \{e^{i \frac{\Phi}{2}} {\Psi}_{+} + e^{-i\frac{\Phi}{2}}{\Psi}_{-}\},$
where $\Phi\in R ~\mbox{mod} ~2\pi$ is the self-adjoint extension parameter. 

Using the properties of the confluent hypergeometric functions, 
at $r\rightarrow0$ we have,
\begin{equation}\label{sae14}
\Psi_{A_+}= \frac{\pi}{\lambda \sin \pi (1+2\beta)} \left[\frac{(\lambda+\beta+i\alpha)}{\Gamma(1-\beta-i\alpha)\Gamma(1+2\beta)}r^{\beta-\frac{1}{2}}-\left(\frac{2}{l}\right)^{-2\beta}\frac{(\lambda-\beta+i\alpha)}{\Gamma(1+\beta-i\alpha)\Gamma(1-2\beta)}r^{-\beta-\frac{1}{2}}\right]
\end{equation}
and
\begin{equation}\label{sae15}
\Psi_{A_-}= \frac{\pi}{\lambda \sin \pi (1+2\beta)} \left[\frac{(\lambda+\beta+i\alpha)}{\Gamma(-\beta+i\alpha)\Gamma(1+2\beta)}r^{\beta-\frac{1}{2}}-\left(\frac{2}{l}\right)^{-2\beta}\frac{(\lambda-\beta+i\alpha)}{\Gamma(\beta+i\alpha)\Gamma(1-2\beta)}r^{-\beta-\frac{1}{2}}\right].
\end{equation}
\begin{figure}
[ht] 
\centering
\includegraphics[bb= 280 14 12 220]{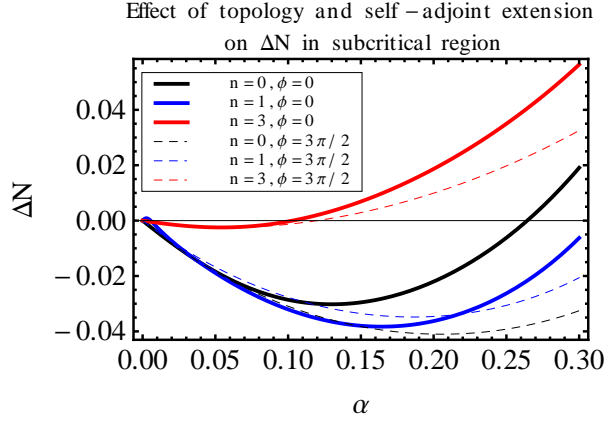}
 % f1.eps: 0x0 pixel, 0dpi, 0.00x0.00 cm, bb=240 14 212 40
\caption{Dependence of $\Delta N$ on subcritical Coulomb potential using 
Friedel sum rule and generalized boundary condition. The solid lines
correspond to $\Phi = 0$ and the dashed lines correspond to $\Phi = \frac{3\pi}{2}$.}
\label{fig:2}
\end{figure}
To find out the scattering phase shift using the generalized boundary conditions,
we first try to obtain the solution of Eq.(\ref{f14}) which gives the physical 
scattering states. The required solution is
\begin{eqnarray}
 \label{sae16}
v(s) = C_1 M \left ( \beta + i\alpha,~ 1 + 2\beta,~s \right) + C_2 s^{-2\beta} M \left(-\beta + i\alpha,~1 - 2\beta,~s \right),
\end{eqnarray}
where $s=-2ikr$.

Now with the help of Eq.(\ref{f12}) we have 
\begin{eqnarray}
 \label{sae17}
u(s) = C_1 \frac{(\beta+i\alpha)}{\lambda} M \left ( 1+\beta + i\alpha,~ 1 + 2\beta,~s \right) + C_2 s^{-2\beta} \frac{(-\beta+i\alpha)}{\lambda} M \left(1-\beta + i\alpha,~1 - 2\beta,~s \right).
\end{eqnarray}
Using Eq.(\ref{sae16}) and Eq.(\ref{sae17}) we get the upper component of the 
wave-function $\Psi$ as
\begin{eqnarray}
 \label{sae18}
\Psi_A (s) &=& r^{\beta-\frac{1}{2}} e^{ikr} \{C_1 \frac{(\beta+i\alpha)}{\lambda} M \left ( 1+\beta + i\alpha,~ 1 + 2\beta,~s \right) \nonumber\\
           &+& C_2 s^{-2\beta} \frac{(-\beta+i\alpha)}{\lambda} M \left(1-\beta + i\alpha,~1 - 2\beta,~s \right)+C_1 M \left ( \beta + i\alpha,~ 1 + 2\beta,~s \right)\nonumber\\
           &+&C_2 s^{-2\beta} M \left(-\beta + i\alpha,~1 - 2\beta,~s \right) \}.
\end{eqnarray}
In the limit $r\rightarrow0$ we match the behaviour of this physical wave function
with a typical element of ${\mathcal{D}}_{\Phi}(H_r)$ to ensure the unitary evolution 
of $H_r$. When $r\rightarrow0$, Eq.(\ref{sae18}) gives
\begin{eqnarray}
 \label{sae19}
\Psi_A (r)=C_1\frac{\lambda+\beta+i\alpha}{\lambda} r^{\beta-\frac{1}{2}} + C_2\frac{(\lambda-\beta+i\alpha)}{\lambda}(-2ik)^{-2\beta}r^{-\beta-\frac{1}{2}}.
\end{eqnarray}
In the same limit a typical element of the domain 
${\mathcal{D}}_{\Phi}(H_r)$ is given by 
\begin{eqnarray}
 \label{sae20}
\Psi (r) = \eta(e^{\frac{i\Phi}{2}}\Psi_+ + e^{\frac{-i\Phi}{2}}\Psi_-).
\end{eqnarray}
Comparing Eq.(\ref{sae19}) and Eq.(\ref{sae20}) we have
\begin{eqnarray}
 \label{sae21}
C_1=\frac{\eta \pi}{\sin\pi(1+2\beta)}\left[\frac{e^{\frac{i\Phi}{2}}}{\Gamma(1-\beta-i\alpha)\Gamma(1+2\beta)} + \frac{e^{\frac{-i\Phi}{2}}}{\Gamma(-\beta+i\alpha)\Gamma(1+2\beta)} \right]
\end{eqnarray}
and
\begin{eqnarray}
 \label{sae22}
C_2=-(-ikl)^{2\beta}\frac{\eta \pi}{\sin\pi(1+2\beta)}\left[\frac{e^{\frac{i\Phi}{2}}}{\Gamma(1+\beta-i\alpha)\Gamma(1-2\beta)} + \frac{e^{\frac{-i\Phi}{2}}}{\Gamma(\beta+i\alpha)\Gamma(1-2\beta)} \right].
\end{eqnarray}
Now to find the scattering matrix and the phase shift we investigate the 
asymptotic behaviour of the wave-function $\Psi$ with the help of the 
properties of the confluent hypergeometric functions. When $r\rightarrow\infty$
we note that
\begin{eqnarray}
 \label{sae23}
\Psi_A (r)=&&\!\!\!\!\!\!\!\!\!\!(-2ik)^{-\beta} (-i)^{i\alpha}\left[C_1 \frac{\beta+i\alpha}{\lambda} \frac{\Gamma(1+2\beta)}{\Gamma(1+\beta+i\alpha)} + C_2 \frac{-\beta+i\alpha}{\lambda}\frac{\Gamma(1-2\beta)}{\Gamma(1-\beta+i\alpha)}\right]\frac{e^{-i[kr-\alpha \ln(2kr)]}}{\sqrt{r}}\nonumber\\
&&\!\!\!\!\!\!\!\!\!\!\!\!\!\!\!\!\!\! +\ (-2ik)^{-\beta} (-i)^{-i\alpha}\left[C_1\frac{\Gamma(1+2\beta)}{\Gamma(1+\beta-i\alpha)}e^{-i\pi(\beta+i\alpha)} + C_2 \frac{\Gamma(1-2\beta)}{\Gamma(1-\beta-i\alpha)}e^{-i\pi(-\beta+i\alpha)}\right]\frac{e^{i[kr-\alpha \ln(2kr)]}}{\sqrt{r}}. \nonumber \\
\end{eqnarray}
The scattering matrix $S$ and the corresponding phase shift $\delta(k)$ is given by
\begin{eqnarray}
\label{sae24}
S=e^{2i\delta(k)}=-\lambda e^{-2i\alpha \ln(2kr)}\left[\frac{F + G}{\lambda^2 F^* + G^*}\right],
\end{eqnarray}
where
\begin{eqnarray}
\label{sae25}
F= \left\{\frac{e^{-\frac{i\pi\beta}{2}}-(kl)^{2\beta}e^{\frac{i\pi\beta}{2}}}{\Gamma(1+\beta-i\alpha)\Gamma(1-\beta-i\alpha)}\right\}e^{\frac{i\Phi}{2}}
\end{eqnarray}
and
\begin{eqnarray}
\label{sae26}
G= -\beta\left\{\frac{e^{-\frac{i\pi\beta}{2}}}{\Gamma(1+\beta-i\alpha)\Gamma(1-\beta+i\alpha)}+\frac{(kl)^{2\beta}e^{\frac{i\pi\beta}{2}}}{\Gamma(1-\beta-i\alpha)\Gamma(1+\beta+i\alpha)}\right\}e^{-\frac{i\Phi}{2}}.
\end{eqnarray}
From Eq.(\ref{sae24}) we can see that the scattering matrix and the phase shift 
explicitly depend on the self-adjoint extension parameter $\Phi$ or equivalently the 
generalized boundary conditions. We should always keep in mind that the conditions are 
valid only for the range $0<\beta<\frac{1}{2}$. Different values of $\Phi$ corresponds to different
combinations of the short range interactions induced by the external Coulomb charge
which gives rise to inequivalent quantum description of the gapless graphene cone.
The value of $\Phi$ cannot be determined analytically but by measuring quantities 
depending on the scattering data, it can be fixed empirically.

The dependence of the change in the number of states on the Coulomb potential
around the Coulomb charge can be determined with the help of the Friedel sum rule,
which connects the scattering phase shifts with the change in the number of states.
Therefore using Eq.(\ref{sae24}) we plot the dependence of the change in the number of
states on Coulomb potential for the parameter range $0<\beta<\frac{1}{2}$ for
different values of $\Phi$.  
From the plot we can clearly see that $\Delta N$ depends on the topology
of the system as well as the boundary condition applied on it.  
 
\section{supercritical region}

In this section for any given value of $n$ and $j$, we always choose $\alpha$ greater than the corresponding 
value of $|\lambda|$ to ensure that the coupling is in the supercritical region. We define $\beta$ as $i\eta$ where 
$\eta=\sqrt{\alpha^2 - \lambda^2}$. Then the solution of Eq. (\ref{f14}) is given by
\begin{equation} \label{f18}
 v^{(j)}(s) = C_1 M \left ( i(\eta + \alpha),~ 1 + 2i\eta,~s \right)
    + C_2 {s}^{- 2 i\eta} M \left (i(\alpha-\eta),~ 1-2i\eta,~s \right).
\end{equation}
From Eqs. (\ref{f12}) and (\ref{f14}) we get
\begin{equation} \label{f19}
 u^{(j)}(s) = -iC_1 \chi M \left (1+i(\eta + \alpha),~ 1 + 2i\eta,~s \right)
    -i (C_2/\chi) {s}^{- 2 i\eta} M \left (1+i(\alpha-\eta),~ 1-2i\eta,~s \right),
\end{equation}
where $\chi=\sqrt{\frac{\alpha+\eta}{\alpha-\eta}}$.
\begin{figure}
[ht] 
\centering
\includegraphics[bb= 280 14 12 220]{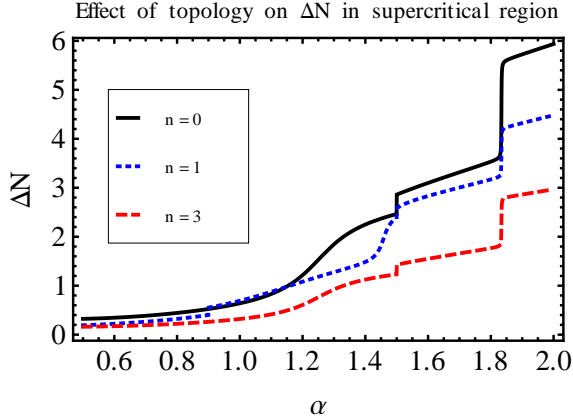}
 % f1.eps: 0x0 pixel, 0dpi, 0.00x0.00 cm, bb=240 14 212 40
\caption{Dependence of change in the number of states on supercritical Coulomb potential using Friedel sum rule.}
\label{fig:3}
\end{figure}

In order to proceed, we use the zigzag edge boundary condition $[u^{(j)}(l_0) - v^{(j)}(l_0)]= 0$, where $l_0$ is a distance from the apex, 
of the order of the lattice scale in graphene.  This gives
\begin{eqnarray}
\label{f20}
C_2 = e^{2i\xi(k)} \chi e^{\pi \eta} C_1~~~~~~\mbox{where}~~~~~e^{2i\xi(k)} = \frac{i(1+i\chi)}{(1-i\chi)}e^{2i\eta \mbox{ln}(2kl_0)}.
\end{eqnarray}
From the above, we obtain the scattering matrix $S$ as 
\begin{eqnarray}
\label{f21}
S = e^{2i\delta_j (k)} =  \left [ \frac{h_{\alpha,\eta}+e^{2i\xi(k)}e^{-\pi \eta}\chi h_{\alpha,-\eta}}{e^{\pi \eta}\chi h^{*}_{\alpha,-\eta}+e^{2i\xi(k)}h^{*}_{\alpha,\eta}} \right ]
e^{-2i\alpha \mbox{ln}(2kr)},
\end{eqnarray}
where $h_{\alpha,\eta}=\frac{\Gamma(1+2i\eta)}{\Gamma(1+i\eta-i\alpha)}$. 

From Eq. (\ref{f21}) we obtain the scattering phase as 
\begin{eqnarray}
\label{f22}
\delta_j (k) = \mbox{arg}[e^{-i\xi(k)} + c e^{i\xi(k)}] - \alpha \mbox{ln}(2kr) + \mbox{arg}(h_{\alpha,\eta}),
\end{eqnarray}
where $c=e^{-\pi \eta}\chi \frac{h_{\alpha,-\eta}}{h_{\alpha,\eta}}$.

From Eq.(\ref{f22}) we can see that the second term $- \alpha \mbox{ln}(2kr)$ in the R.H.S.
is also present in the subcritical region. It is typical for a phase coming from the Coulomb tail
and does not affect the polarization at a finite distance\cite{levi1}. In addition to that term
the scattering phase shift in the supercritical region has a strong energy dependence
through the first term in the R.H.S. of Eq.(\ref{f22}). Keeping in mind the relation between 
polarization charge and change in the number of states we can find out the 
dependence of the polarization charge on the supercritical Coulomb potential from the 
scattering phase shift at Fermi energy according to the Friedel sum rule using Eq.(\ref{f22}).

Though the nature of dependence of $\Delta N$ on supercritical Coulomb potential 
is quite different from that of the subcritical region, here also we can see that for  different values of $n$, 
$\Delta N$ increases with the supercritical Coulomb potential in a different manner.
The sharp increase in $\Delta N$ at certain values of $\alpha$ corresponds 
to the quasibound states formed in this region.
The plot shows that the rate of change in the number of states 
around the external supercritical Coulomb charge
changes with the angular deficit of the graphene cone.

\section{Conclusion}
In this paper we have studied the Friedel sum rule for graphene 
with a conical defect in presence of an external Coulomb charge.
The eigenstates of the Dirac equation 
valid for the low energy excitations in graphene cone\cite{critical1}
have been used to obtain the relation between the change in the 
number of states $\Delta N$ due to the Coulomb impurity and the summation 
of the scattering phase shifts at Fermi energy in different angular 
momentum channels. As the scattering phase shifts explicitly depend on the topology of 
the graphene cone, we have shown $\Delta N$ will also depend on the angle of the cone.

We have plotted the dependence of $\Delta N$ on Coulomb potential
using the Friedel sum rule for both subcritical and supercritical
values of the Coulomb impurity. In the subcritical region from Fig.\ref{fig:1}
we can see that $\Delta N$ increases with the increase in the value of $n$
for a certain value of $\alpha$. As polarization 
charge is directly proportional to $\Delta N$, we can say that for a fixed 
value of external subcritical Coulomb charge, polarization charge increases with the decrease
in opening angle of the graphene cone. The conical defect and the external charge impurity 
can lead to short range interactions in graphene.
Those interactions cannot be directly included in the Dirac equation because the 
latter is valid only in the long wavelength limit\cite{reed,falomir,critical2,ksg1}. 
The single real parameter which labels the boundary conditions can be thought of as 
encoding the combined effect of all those short range interactions. This
parameter is also necessary for ensuring conservation of probability current and unitary
time evolution of the system\cite{dipti1,dipti2,dipti3}. 
The scattering phase shifts and $\Delta N$
depend on this parameter explicitly within the specified system 
parameter range. The parameter can only be determined empirically 
as theory cannot predict its value.

The analysis for the supercritical region\cite{critical1,castro,levi1,levi2,kats1,us1,levi3}
has been done using the zigzag edge boundary condition. The sharp increase in $\Delta N$
at certain values of $\alpha$ corresponds to the quasibound states
formed in that region. Recently such quasibound states have been observed
experimentally for plane massless graphene\cite{levi3}. Here the analysis
has been done for massless graphene in presence of a conical defect. Fig.\ref{fig:3} shows that as the 
value of $n$ increases, $\Delta N$ decreases for a specific value of $\alpha$. 
Therefore for supercritical region we can say that for a fixed 
value of external Coulomb charge, polarization charge decreases with the 
decrease in opening angle of the graphene cone.

In this paper we have considered only the gapless excitations
of a graphene cone. A similar analysis for the gapped excitations
can also be interesting which is currently under consideration.
%\section*{References}


\begin{thebibliography}{99}

\bibitem{friedel}J. Friedel, Philos. Mag. {\bf 43}, 153 (1952).
\bibitem{mahan} G. D. Mahan, Many-Particle Physics (Plenum, New York, 2000), p. 195.

\bibitem{lin1} D.-H. Lin, Phys. Rev. {\bf A 72}, 012701 (2005).
\bibitem{lin2} D.-H. Lin, Phys. Rev. {\bf A 73}, 052113 (2006).
\bibitem{lin3} D.-H. Lin, J.M.P. {\bf 47}, 042302 (2006).

\bibitem{moroz1} A. Moroz, Phys. Lett. {\bf B 358}, 305 (1995).
\bibitem{moroz2} A. Moroz, Phys. Rev. {\bf A 53}, 669 (1996).

\bibitem{critical1} B Chakraborty, Kumar S. Gupta and S. Sen, Phys. Rev. {\bf B 83} 115412 (2011).

\bibitem{castro} V. M. Pereira, J. Nilsson and A. H. Castro Neto, Phys. Rev. Lett. {\bf 99}, 166802 (2007).
\bibitem{levi1} A. V. Shytov, M. I. Katsnelson and L. S. Levitov, Phys. Rev. Lett. {\bf 99}, 236801 (2007).
\bibitem{levi2} A. V. Shytov, M. I. Katsnelson and L. S. Levitov, Phys. Rev. Lett. {\bf 99}, 246802 (2007).
\bibitem{kats1} A. Shytov, M. Rudner, N. Gu, M. Katsnelson and L. Levitov, Solid State Comm. {\bf 149}, 1087 (2009).
\bibitem{us1} Kumar S. Gupta and Siddhartha Sen, Mod. Phys. Lett. {\bf A 24}, 99 (2009).


\bibitem{novo1} Novoselov K S, Geim A K, Morozov S V, Jiang D, Katsnelson M I, Grigorieva I V and Firsov A A 2004 {\it Science} {\bf 306} 666. 
\bibitem{novo2} Novoselov K S, Geim A K, Morozov S V, Jiang D, Katsnelson M I, Grigorieva I V, Dubonos S V and Firsov A A 2005 {\it Nature} {\bf 438} 197. 
\bibitem{zhang} Zhang Y, Tan Y-W, Stormer H L and Kim P 2005 {\it Nature} {\bf 438} 201. 

\bibitem{levi3} Wang Yang et al. Science {\bf 340}, 734 (2013).

\bibitem{wall} Wallace P R 1947  The band theory of graphite. {\it Phys. Rev.} {\bf 71} 622. 
\bibitem{mele} DiVincenzo D P and Mele E J 1984 {\it Phys. Rev.} {\bf B 29} 1685. 
\bibitem{sem} Semenoff G W 1984 {\it Phys. Rev. Lett.} {\bf 53} 2449. 
\bibitem{geim} Geim A K and Novoselov K S 2007 {\it Nature Materials} {\bf 6} 183. 

\bibitem{rmp1} Castro Neto A H, Guinea F, Peres N M R, Novoselov K S and Geim A K 2009 {\it Rev. Mod. Phys.} {\bf 81} 109. 
\bibitem{rmp2} Peres N M R 2010 {\it Rev. Mod. Phys.} {\bf 82} 2673.  
\bibitem{rmp3} Das Sarma S, Adam S, Hwang E H and Rossi E 2011 {\it Rev. Mod. Phys.} {\bf 83} 407. 

\bibitem{crespi1} Lammert P E, Crespi V H 2000 {\it Phys. Rev. Lett.} {\bf 85} 5190. 
\bibitem{crespi2} Lammert P E, Crespi V H 2004 {\it Phys. Rev.} {\bf B 69} 035406. 
\bibitem{stone1} Pachos J K, Stone M and Temme K 2008 {\it Phys. Rev. Lett.} {\bf100} 156806. 

\bibitem{voz1} Gonzalez J, Guinea F and Vozmediano M A H 1992 {\it Phys. Rev. Lett.} {\bf 69} 172. 
\bibitem{voz2} Gonzalez J, Guinea F and Vozmediano M A H 1993 {\it Nucl. Phys.}{\bf B 406} 771. 

\bibitem{osi1} Kolesnikov D V and Osipov V A 2006 {\it Eur. Phys. J.} {\bf B 49} 465. 
\bibitem{sitenko} Sitenko Y A  and Vlasii N D 2007 {\it Nuclear Physics} {\bf B 787} 241.

\bibitem{voz3} Cortijo A and Vozmediano M A H 2007 {\it Nucl. Phys.}{\bf B 763} 293. 
\bibitem{voz4} de Juan F, Cortijo A and Vozmediano M A H 2007 {\it Phys. Rev.} {\bf B 76} 165409. 

\bibitem{mudry} Hou C Y, Chamon C and Mudry C 2007 {\it Phys. Rev. Lett.} {\bf98} 186809. 

\bibitem{furtado} Furtado C, Moraes F, Carvalho A M de M 2008 {\it Phys. Lett.} {\bf A 372} 5368. 
\bibitem{stone2} Roy A and Stone M 2010 {\it J. Phys.} {\bf A 43} 015203. 
\bibitem{voz5} Vozmediano M A H, Katsnelson M I and Guinea F 2010 {\it Phys. Reports} {\bf 496} 109-148. 
\bibitem{Gonzalez} Gonzalez J and Herrero J 2010 {\it Nucl. Phys.} {\bf B 825} 426-443. 
\bibitem{yazyev} Yazyev O V and Louie S G 2010 {\it Phys. Rev.} {\bf B 81} 195420. 
\bibitem{fonseca} Fonseca J M, Moura-Melo W A and Pereira A R 2010 {\it Phys. Lett.} {\bf A 374} 4359. 
\bibitem{voz6} de Juan F, Cortijo A, Vozmediano M A H and Cano A 2011 {\it Nature Physics} {\bf 7} 810. 
\bibitem{Guinea1} Abedpour N, Asgari R and Guinea F 2011 {\it Phys. Rev.} {\bf B 84} 115437. 
\bibitem{furtado2} Bakke K, Petrov A Y and Furtado C 2012 {\it Annals Phys.} {\bf 327}2946.
\bibitem{voz7} Cortijo A, Guinea F and Vozmediano M A H 2012 {\it J. Phys.} {\bf A 45} 383001.

\bibitem{Jackiw1} de Sousa Gerbert P and Jackiw R 1989 {\it Commun. Math. Phys.} {\bf 124} 229-260 
\bibitem{gerbert} de Sousa Gerbert P 1989 {\it Phys. rev.} {\bf D 40} 1346-49 
\bibitem{yam} Yamagishi H 1983 {\it Phys. Rev.} {\bf D 27} 2383 
\bibitem{Jackiw2} Jackiw R and Pi S Y 2007, {\it Phys. Rev. Lett.} {\bf 98} 266402 
\bibitem{Jackiw3} Chamon C, Hou C Y, Jackiw R, Mudry C, Pi S Y and Semenoff G 2008 {\it Phys.Rev.} {\bf B 77} 235431 
\bibitem{Jackiw4} Jackiw R and Pi S Y 2008 {\it Phys. Rev.} {\bf B 78} 132104 
\bibitem{Jackiw5} Jackiw R, Milstein A I, Pi S Y and Terekhov I S 2009 {\it Phys. Rev.} {\bf B 80} 033413 


\bibitem{reed} Reed M and Simon B 1972{\it Methods of Modern Mathematical Physics},volume 2, (Academic Press, New York)
\bibitem{falomir} Falomir H and Pisani P A G 2001 {\it J. Phys. A : Math. Gen.} {\bf 34} 4143 
\bibitem{critical2} Kumar S. Gupta and S. Sen, Phys. Rev. {\bf B 78}, 205429 (2008).
\bibitem{ksg1} Gupta K S, Samsarov A and Sen S 2010 {\it Eur. Phys. J.} {\bf B 73} 389 

\bibitem{dipti1} Sen Diptiman and Deb Oindrila, Phys. Rev. {\bf B 85}, 245402 (2012). 
\bibitem{dipti2} Soori Abhiram, Deb Oindrila, Sengupta K. and Sen Diptiman, Phys. Rev. {\bf B 87}, 245435 (2013).
\bibitem{dipti3} Deb Oindrila, Soori Abhiram and Sen Diptiman,  arXiv:1401.1027 (2014). 

\bibitem{stegun} M. Abramowitz and I.A. Stegun, {\it Handbook of Mathematical Functions} (Dover, New York, 1970).

\end{thebibliography}
\end{document}